\documentclass[sigconf, nonacm]{acmart}
\AtBeginDocument{%
  \providecommand\BibTeX{{%
    \normalfont B\kern-0.5em{\scshape i\kern-0.25em b}\kern-0.8em\TeX}}}

\begin{document}

\title{SSE: A Metric for Evaluating Search System Explainability}

\author{Catherine Chen}
\email{catherine_s_chen@brown.edu}
\affiliation{
  \institution{Brown University}
  \city{Providence}
  \state{Rhode Island}
  \country{USA}
}

\author{Carsten Eickhoff}
\email{c.eickhoff@acm.org}
\affiliation{
  \institution{University of T\"{u}bingen}
  \city{T\"{u}bingen}
  \country{Germany}
}

% \renewcommand{\shortauthors}{Chen and Eickhoff}

\begin{abstract}
Explainable Information Retrieval (XIR) is a growing research area focused on enhancing transparency and trustworthiness of the complex decision-making processes taking place in modern information retrieval systems. While there has been progress in developing XIR systems, empirical evaluation tools to assess the degree of explainability attained by such systems are lacking. To close this gap and gain insights into the true merit of XIR systems, we extend existing insights from a factor analysis of search explainability to introduce SSE (Search System Explainability), an evaluation metric for XIR search systems. Through a crowdsourced user study, we demonstrate SSE's ability to distinguish between explainable and non-explainable systems, showing that systems with higher scores indeed indicate greater interpretability. Additionally, we observe comparable perceived temporal demand and performance levels between non-native and native English speakers. We hope that aside from these concrete contributions to XIR, this line of work will serve as a blueprint for similar explainability evaluation efforts in other domains of machine learning and natural language processing.
\end{abstract}




\maketitle

\section{Introduction}

Explainable information retrieval (XIR) research aims to develop methods that increase the transparency and reliability of information retrieval systems. XIR systems are designed to provide end-users with a deeper understanding of the rationale behind result ranking decisions. Besides casual Web search, these systems hold promising potential for impactful real-world information needs, such as matching patients to clinical trials, retrieving case law for legal research, and detecting misinformation in news and media analysis. Despite several advancements in their development, there is a lack of empirical, standardized techniques for evaluating the efficacy of XIR systems. 

Current approaches toward evaluating XIR systems are limited by a lack of consensus in the broader explainable artificial intelligence (XAI) community on a definition of explainability. Explainability has often been considered a singular concept although recent literature suggests it to be an amalgamation of several sub-factors \cite{lipton2018mythos, doshi2018considerations, nauta2022anecdotal}. As a result, evaluation has occurred on a binary scale, considering systems as either explainable or black-box, hindering direct system comparison. Furthermore, explainability is often declared with anecdotal evidence rather than measured quantitatively. 

\citet{chen2022evaluating} take a step towards addressing these limitations by introducing a multidimensional definition of explainability in search systems. In this paper, we extend their two-factor hierarchical definition to propose a continuous-scale evaluation metric for quantifying explainability in Web search systems. To validate this metric and gain practical insights into implementing human-centered evaluation tools, we conduct a crowdsourced user study and assess the impact such tools have on human annotators.\footnote{ We make all data and analysis code publicly available at \textit{https://github.com/catherineschen/sse-metric}.} Specifically, we make the following contributions: 

\begin{itemize}
    \item We introduce \textit{SSE}\footnote{Pronounced `es-es-e'}, a metric to evaluate \textit{Search System Explainability}, extending the search system explainability factor analysis by \citet{chen2022evaluating}.
    \item We conduct a crowdsourced user study and validate the ability of SSE to distinguish between non-explainable and explainable systems, demonstrating that higher SSE scores indicate greater explainability. 
    \item We evaluate the questionnaire workload on crowdworkers and find that while non-native English speakers report slightly higher levels of perceived \textit{Mental Demand} and \textit{Effort}, their perceived \textit{Temporal Demand} and \textit{Performance} are comparable to those of native English speakers.
\end{itemize}

The remainder of the paper is structured as follows: In Section \ref{related-work}, we present related work in XIR system evaluation. Section \ref{metric-overview} provides an overview search system explainability evaluation. In Section \ref{study-design}, we detail our study design and data collection setup. In Section \ref{results}, we discuss the evaluation results of SSE, examining its effectiveness in evaluating search system explainability and the workload it imposes on crowdworkers. Finally, in Section \ref{conclusion}, we conclude with an overview of future research directions.

\section{Related Work} \label{related-work}

Due to the lack of consensus and recognition of explainability as a multi-facet concept, evaluation falls short in two ways. Firstly, current methods do not consider a multidimensional definition. Authors of frameworks such as LIME and SHAP demonstrate the utility of their methods through user evaluations but fail to quantify the degree of explainability provided by their methods \citep{ribeiro2016should, lundberg2017unified}. Relying solely on declarations of explainability without measurement hinders targeted system improvements and a more holistic definition is needed for robustness.

Secondly, while there are many ML evaluation metrics for system performance comparison, there is no standard metric for evaluating explainability. Although several authors acknowledge the importance of quantitative evaluation metrics \citep{nauta2022anecdotal, doshi2017towards, adadi2018peeking, das2020opportunities}, existing approaches often focus on a singular aspect of explainability or lack evaluation of explanation quality, relying on anecdotal evidence \citep{singh2019exs, qu2020towards, yu2022towards, polley2021towards, colin2022cannot}. While \citet{nguyen2020quantitative} come close by introducing a suite of metrics to quantify interpretability, they fail to quantify interactions between facets and measure the importance of each individual facet. 

\citet{chen2022evaluating} address the first limitation by introducing an empirically derived multidimensional definition of explainability through factor analysis. In this paper, we address the second limitation by extending the insights of their work and introducing an all-encompassing metric that takes into account multiple desiderata of explainability, which will help us gain more insight into the strengths and weaknesses of a system.

\section{Search System Explainability (SSE)} \label{metric-overview}

The search explainability factor model introduced by \citet{chen2022evaluating} consists of two factor groups, \textit{utility} and \textit{roadblocks}, where the first factor broadly refers to a system's usefulness within some context and the second factor refers to qualities a system lacks in order to be fully explainable. The factor model originates from a 19-item questionnaire (e.g., \textit{``I'm unable to follow how the search engine ordered the results.''}) scored on a 7-point Likert scale (1 - Strongly Disagree, 7 - Strongly Agree). Administered following user interaction with a specific search system (e.g., via a search task), the questionnaire assesses the extent to which system features facilitate or impede explainability.

    \begin{equation} \label{eq: score}
        SSE = MinMax Norm \bigg( \sum_{f\in{F}} w_f \sum_{i\in{I_f}} w_i r_i \bigg)
    \end{equation}

We extend this factor analysis to introduce Search System Explainability (SSE), a metric for evaluating search system explainability. The formula for SSE is shown in Equation \ref{eq: score}, where $F$ is the set of all factors $f$ and $I_f$ is the set of survey items $i$ that correspond to factor $f$. The user's response $r_i$ to item $i$ is weighted by the corresponding loading coefficient $w_i$  and the accumulated response scores for each factor are weighted by the corresponding loading coefficient $w_f$ determined by \citet{chen2022evaluating}. To make SSE score values more interpretable, min-max normalization with the smallest and largest theoretically possible SSE scores is applied to normalize overall values between 0 and 1.

\section{Study Design} \label{study-design}

In this paper, we aim to test the hypothesis that an explainable system will attain a higher SSE score than a non-explainable system.

\subsection{Task Setup}

\begin{figure}[h!]
    \centering
    \includegraphics[width=0.22\textwidth]{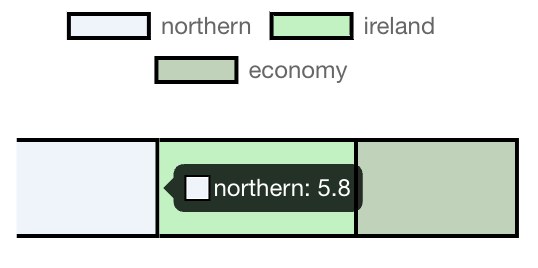}
    \caption{\texttt{BARS} search system based on the system presented by \citet{ramos2020search}. Stacked bar graphs, located alongside each search result, depict the BM25 scores assigned to individual keywords in the search query. Hovering over a bar reveals the corresponding score for the respective term.}
    \label{fig:bars}
\end{figure}

We employed a between-subjects design and asked participants to evaluate one of two search systems: (A) \texttt{BASELINE}: a minimalist search engine resembling existing commercial search systems with no explainable capabilities and (B) \texttt{BARS}: a search system with visual explanations representing query term importance (Figure \ref{fig:bars}) developed by ~\citet{ramos2020search}. To guide their evaluation, we provided users access to a collection of documents and asked them to answer an open-ended question on one of 10 possible topics\footnote{List of topics: compost pile, Antarctica exploration, heroic acts, Northern Ireland industry, new hydroelectric projects, health and computer terminals, quilts and income, wildlife extinction, illegal technology transfer, Salvation Army benefits} using the search engine. Documents, questions, and topics were selected from the TREC 2004 Robust Track Dataset \cite{voorhees_overview_2005}.

Both search systems utilized the same BM25 retrieval model to rank relevant documents. We followed the document selection procedure described in \citet{chen2022evaluating} to sample 50 relevant documents from each of the 10 topics, resulting in a collection of 500 documents for users to search from. We used PyTerrier \cite{pyterrier2020ictir} to index the documents and wrapped the index in a Flask server, hosting it on Fly.io. We hosted the front-end search system on Netlify and used MongoDB for additional data collection.

After completing the search task, participants were asked to fill an evaluation questionnaire. We incorporated two attention check items (``\textit{I swim across the Atlantic Ocean to get to work every day.}'' and ``\textit{I think search engines are cool. Regardless of your opinion, select `Strongly Agree' as your response below. This is an attention check.}'') at random points in the questionnaire for quality control.

Following the evaluation questionnaire, we administered the NASA Task Load Index (NASA-TLX) \cite{hart2006nasa} to measure the perceived workload of the evaluation questionnaire. The NASA-TLX is a 6-item questionnaire measuring perceived workload on 6 dimensions (\textit{Mental Demand, Physical Demand, Temporal Demand, Performance, Effort, Frustration}) with responses recorded on a scale of 0-20, where lower scores are more desirable and indicate a low workload (except for Performance, where lower scores indicate high success). Since the original scoring method involves a two-step weighting scheme through numerous pairwise comparisons, we followed the common practice of analyzing raw TLX (RTLX) responses to reduce the amount of time needed to administer the survey \cite{hart2006nasa}.

\subsection{Data Collection}

\begin{figure*}[h!]
    \centering
    \includegraphics[width=0.70\textwidth]{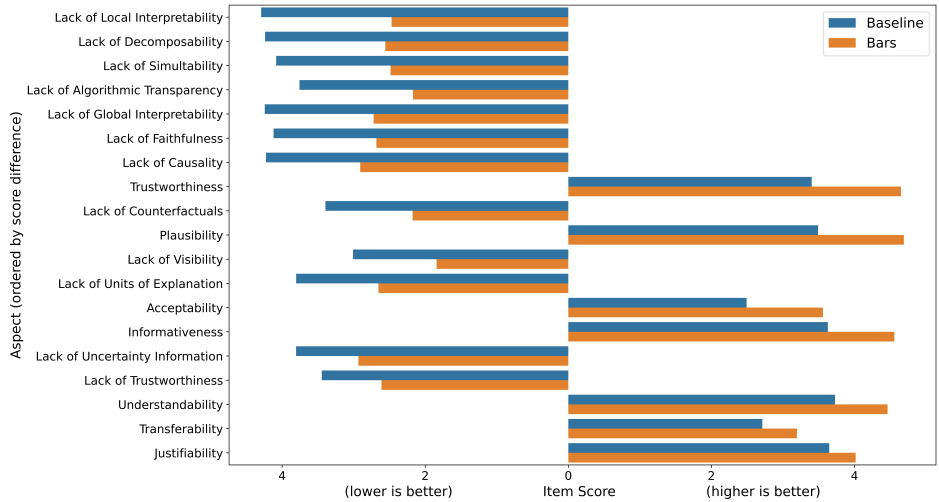}
    \caption{Loadings for individual questionnaire items. Items are labeled by their original aspect as determined by \citet{chen2022evaluating} and ordered by score (response multiplied by item loading) difference between the two systems. Aspects with bars increasing toward the right indicate the practical usefulness of each system, while aspects with bars increasing toward the left represent areas requiring improvement to achieve full explainability.}
    \label{fig:aspect-scores}
\end{figure*}

Prior to full-scale data collection, we conducted a think-aloud study with 6 graduate students to identify and address bugs in the study design and system. We used the concurrent think-aloud method due to its effectiveness in usability testing and time efficiency \cite{olmsted2010think}. Observations from this period led to several changes. Specifically, we made slight modifications to the NASA-TLX assessment: (1) we replaced the term `task' with `survey' to align responses with the evaluation questionnaire rather than the search task and (2) we inverted the scale on the \textit{Performance} dimension to have higher responses equate to `Perfect' rather than `Failure' and align with a more intuitive understanding of performance. Additionally, we added more detailed descriptions to certain questions to accurately anchor neutral responses of low effort to 0 instead of 10, which would indicate a medium effort.

We recruited 100 participants aged 18-65 for our study on Prolific\footnote{IRB approval was judged unnecessary by an IRB member due to the classification of our study as a system evaluation.} and paid them £5.25 for 30 minutes of work (estimated from a 15-participant pilot study), equivalent to the minimum wage in our location. Additionally, we required participants to be fluent in English and have a minimum prior approval rate of 95\%. On average, users completed the task more quickly than our initial estimate, with the median time being approximately 22 minutes. 15 countries were represented in our sample, with participants from Australia (1), Africa (4), Europe (81), North America (12), South America (1), and 1 unknown location.

\section{Results and Discussion} \label{results}

\subsection{Search Efficiency}

Despite users in the \texttt{BARS} group spending a slightly longer duration on the overall search task (11.97 min) compared to the \texttt{BASELINE} group (11.41 min), an analysis of search time per query and the number of viewed documents per query indicates that the explainable \texttt{BARS} system exhibited higher efficiency than the non-transparent system (Table \ref{tab:stats}). Users issued more queries per task in the explainable system (3.45 vs 2.54). This may be attributed to query modifications aimed at targeting specific keywords to leverage query-term matching for calculating BM25 scores. These results suggest that the \texttt{BARS} system allowed users to scan search engine result pages (SERP) more quickly. This observation aligns with previous research by \citet{ramos2020search}, which demonstrated the positive impact of explainable systems on search efficiency.

\begin{table}[h!]
\caption{Task engagement statistics (M $\pm$ SD). Although participants in the \texttt{BARS} group spent more time overall compared to the \texttt{BASELINE} group, they were more efficient per SERP.}
\begin{tabular}{@{}lrr@{}}
\toprule
                           & \texttt{BASELINE}   & \texttt{BARS}         \\ 
                           & n=56               & n=44 \\
\midrule
Task time (in min)         & 11.41 $\pm$ 8.81  & 11.97 $\pm$ 6.42 \\
\# of queries              & 2.54  $\pm$ 1.95  & 3.45  $\pm$ 2.90 \\
\# of documents viewed     & 6.30  $\pm$ 4.82  & 6.89  $\pm$ 5.56 \\
Time per query             & 6.79  $\pm$ 8.22  & 5.36  $\pm$ 4.50 \\
Documents viewed per query & 3.84  $\pm$ 4.65  & 2.91  $\pm$ 2.85 \\ 
\bottomrule
\end{tabular}
\label{tab:stats}
\end{table}

\begin{table*}[h!]
\caption{NASA-RTLX responses (lower is better) for perceived evaluation questionnaire workload indicate that the \texttt{BARS} system caused less \textit{Mental} and \textit{Physical Demand}, higher perceived \textit{Performance}, and less perceived \textit{Effort} and \textit{Frustration} than the non-transparent \texttt{BASELINE} system.}
\begin{tabular}{lrrrrrr}
\toprule
               & Mental Demand & Physical Demand & Temporal Demand & Performance & Effort & Frustration \\ \midrule
Overall        & 7.53          & 2.26            & 3.44            & 4.12       & 10.01  & 3.28        \\ \midrule
\texttt{BASELINE}       & 7.79          & 2.89            & 3.02            & 4.02       & 10.18  & 4.00        \\
\texttt{BARS}           & 7.20          & 1.45            & 3.98            & 4.24       & 9.80   & 2.36        \\ \midrule
\end{tabular}
\label{tab:tlx-results}
\end{table*}

\subsection{System Explainability}

\begin{figure}[h!]
    \centering
    \includegraphics[width=0.35\textwidth]{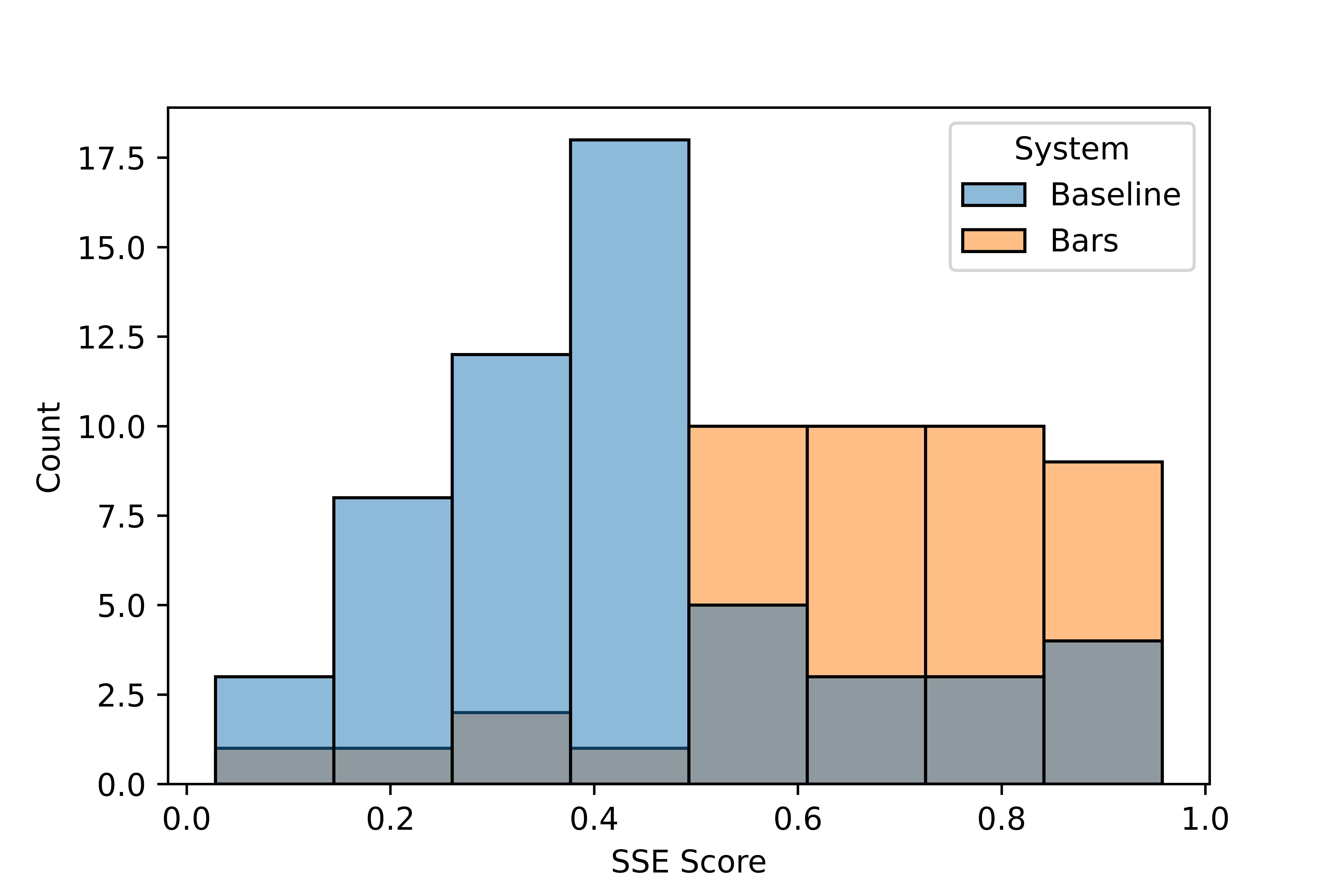}
    \caption{Distribution of SSE scores. Results from a Wilcoxon signed rank test (T = 98.0, p < 0.001) indicate there is a statistically significant difference in scores between systems.}
    \label{fig:scores}
\end{figure}

Congruent with our initial hypothesis, we find that users assigned higher explainability scores to the \texttt{BARS} system in comparison to the non-transparent \texttt{BASELINE} system (Figure \ref{fig:scores}). The \texttt{BARS} system exhibited a mean SSE score (M=0.67) greater than that of the \texttt{BASELINE} system (M=0.44). The results from a Wilcoxon signed rank test on a random sample of the maximum number of samples in the minority class (n=44) reveal a statistically significant difference (T=98.0, p<0.001), confirming that SSE is able to accurately discern between explainable and non-explainable systems.

The aspect scores offer a comprehensive evaluation of individual questionnaire items and specific areas where the \texttt{BARS} system outperformed the \texttt{BASELINE} (Figure \ref{fig:aspect-scores}). Notably, items are grouped by their representation of attributes relating to utility (right) and areas for improvement (left). Additionally, the aspects positioned closer to the top of the graph highlight the most significant disparities in average item scores between the two systems. For instance, the widest margin between the two systems is on \textit{lack of local interpretability}, indicating that the \texttt{BARS} system has more \textit{local interpretability} than the \texttt{BASELINE} system, but still has some room for improvement. Overall, the \texttt{BARS} system outperformed the \texttt{BASELINE} system across all questionnaire items, thus indicating higher readiness for deployment and a reduced need for extensive feature modifications to achieve full explainable capability.

\subsection{Evaluation Questionnaire Task Load}

Table \ref{tab:tlx-results} reports the average NASA-RTLX scores for the 6 workload dimensions. Participants evaluating the \texttt{BARS} system reported a slightly heightened sense of urgency (i.e., \textit{Temporal Demand}), potentially due to the nature of the search task rather than the survey itself, as users in this group took slightly longer to complete the initial search task compared to the \texttt{BASELINE} group (Table \ref{tab:stats}). Overall, participants from both groups considered the evaluation instrument moderately demanding in terms of \textit{Mental Demand}, moderate in \textit{Effort}, low in \textit{Physical Demand}, \textit{Temporal Demand}, and \textit{Frustration}, and high success in \textit{Performance}.

\begin{figure}[h!]
    \centering
    \includegraphics[width=0.35\textwidth]{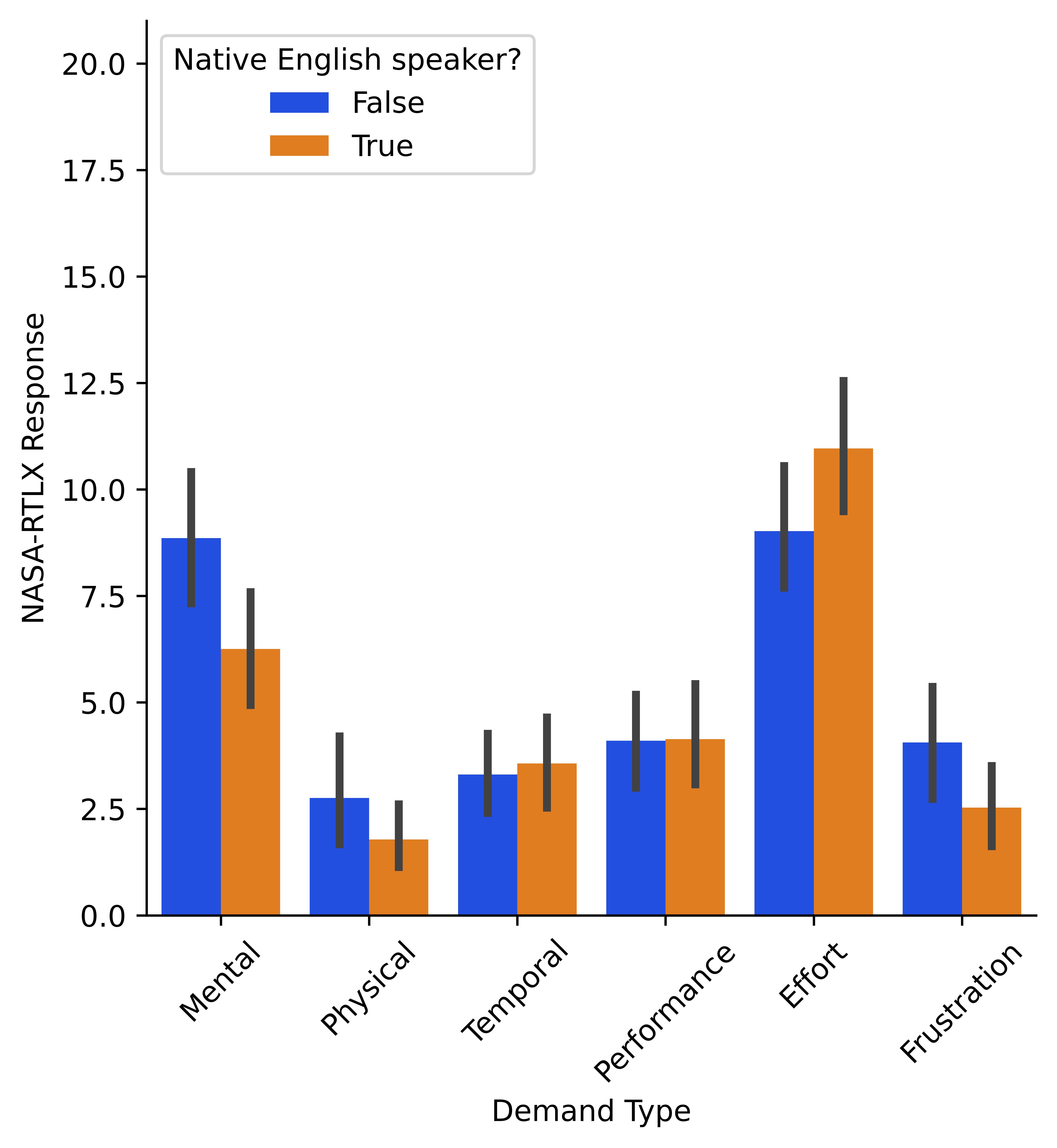}
    \caption{NASA-RTLX responses between native (n=51) and non-native (n=49) English speakers. While non-native English speakers have a higher perceived task load than native English speakers on 3 dimensions, they had a lower perceived \textit{Temporal Demand} and \textit{Effort} for a higher perceived \textit{Performance}.}
    \label{fig:tlx-native}
\end{figure}

Figure \ref{fig:tlx-native} presents the NASA-RTLX scores grouped by native and non-native English speakers. On average, non-native English speakers demonstrated higher perceived levels of \textit{Mental Demand}, \textit{Physical Demand}, and \textit{Frustration} compared to native English speakers. However, their perceived \textit{Performance} was similar to that of native English speakers while reporting less \textit{Temporal Demand} and \textit{Effort}. Nevertheless, these findings are encouraging since the overall task load for all participants ranged from low to medium, suggesting the suitability of the explainability evaluation questionnaire for both native and non-native English speakers.

\section{Conclusion and Future Work} \label{conclusion}

In this paper, we proposed a new metric for evaluating search system explainability on a continuous scale. Through a crowdsourced user study, we demonstrate that an XIR search system exhibited a significantly higher SSE score than a baseline non-transparent search system, indicating greater explainable capability. Additionally, a task load analysis reveals that both native and non-native English speakers found the evaluation instrument to have a low-to-medium workload. It is important to note that explainability can differ between domains and thus, the findings in this paper are limited to Web search. Future work should examine how SSE performs in other contexts. Overall, our study offers valuable insights for future researchers in the field of XIR, providing guidance on utilizing this evaluation instrument to conduct their own assessments.

% \begin{acks}
% \end{acks}

\bibliographystyle{ACM-Reference-Format}
\bibliography{sample-base}


\begin{thebibliography}{20}

%%% the Web address.

\ifx \showCODEN    \undefined \def \showCODEN     #1{\unskip}     \fi
\ifx \showDOI      \undefined \def \showDOI       #1{#1}\fi
\ifx \showISBNx    \undefined \def \showISBNx     #1{\unskip}     \fi
\ifx \showISBNxiii \undefined \def \showISBNxiii  #1{\unskip}     \fi
\ifx \showISSN     \undefined \def \showISSN      #1{\unskip}     \fi
\ifx \showLCCN     \undefined \def \showLCCN      #1{\unskip}     \fi
\ifx \shownote     \undefined \def \shownote      #1{#1}          \fi
\ifx \showarticletitle \undefined \def \showarticletitle #1{#1}   \fi
\ifx \showURL      \undefined \def \showURL       {\relax}        \fi
% invisible to TeX
\providecommand\bibfield[2]{#2}
\providecommand\bibinfo[2]{#2}
\providecommand\natexlab[1]{#1}
\providecommand\showeprint[2][]{arXiv:#2}

\bibitem[\protect\citeauthoryear{Adadi and Berrada}{Adadi and Berrada}{2018}]%
        {adadi2018peeking}
\bibfield{author}{\bibinfo{person}{Amina Adadi} {and} \bibinfo{person}{Mohammed
  Berrada}.} \bibinfo{year}{2018}\natexlab{}.
\newblock \showarticletitle{Peeking inside the black-box: a survey on
  explainable artificial intelligence (XAI)}.
\newblock \bibinfo{journal}{\emph{IEEE access}}  \bibinfo{volume}{6}
  (\bibinfo{year}{2018}), \bibinfo{pages}{52138--52160}.
\newblock


\bibitem[\protect\citeauthoryear{Chen and Eickhoff}{Chen and Eickhoff}{2022}]%
        {chen2022evaluating}
\bibfield{author}{\bibinfo{person}{Catherine Chen} {and}
  \bibinfo{person}{Carsten Eickhoff}.} \bibinfo{year}{2022}\natexlab{}.
\newblock \showarticletitle{Evaluating Search Explainability with Psychometrics
  and Crowdsourcing}.
\newblock \bibinfo{journal}{\emph{arXiv preprint arXiv:2210.09430}}
  (\bibinfo{year}{2022}).
\newblock


\bibitem[\protect\citeauthoryear{Colin, Fel, Cad{\`e}ne, and Serre}{Colin
  et~al\mbox{.}}{2022}]%
        {colin2022cannot}
\bibfield{author}{\bibinfo{person}{Julien Colin}, \bibinfo{person}{Thomas Fel},
  \bibinfo{person}{R{\'e}mi Cad{\`e}ne}, {and} \bibinfo{person}{Thomas Serre}.}
  \bibinfo{year}{2022}\natexlab{}.
\newblock \showarticletitle{What i cannot predict, i do not understand: A
  human-centered evaluation framework for explainability methods}.
\newblock \bibinfo{journal}{\emph{Advances in Neural Information Processing
  Systems}}  \bibinfo{volume}{35} (\bibinfo{year}{2022}),
  \bibinfo{pages}{2832--2845}.
\newblock


\bibitem[\protect\citeauthoryear{Das and Rad}{Das and Rad}{2020}]%
        {das2020opportunities}
\bibfield{author}{\bibinfo{person}{Arun Das} {and} \bibinfo{person}{Paul Rad}.}
  \bibinfo{year}{2020}\natexlab{}.
\newblock \showarticletitle{Opportunities and challenges in explainable
  artificial intelligence (xai): A survey}.
\newblock \bibinfo{journal}{\emph{arXiv preprint arXiv:2006.11371}}
  (\bibinfo{year}{2020}).
\newblock


\bibitem[\protect\citeauthoryear{Doshi-Velez and Kim}{Doshi-Velez and
  Kim}{2017}]%
        {doshi2017towards}
\bibfield{author}{\bibinfo{person}{Finale Doshi-Velez} {and}
  \bibinfo{person}{Been Kim}.} \bibinfo{year}{2017}\natexlab{}.
\newblock \showarticletitle{Towards a rigorous science of interpretable machine
  learning}.
\newblock \bibinfo{journal}{\emph{arXiv preprint arXiv:1702.08608}}
  (\bibinfo{year}{2017}).
\newblock


\bibitem[\protect\citeauthoryear{Doshi-Velez and Kim}{Doshi-Velez and
  Kim}{2018}]%
        {doshi2018considerations}
\bibfield{author}{\bibinfo{person}{Finale Doshi-Velez} {and}
  \bibinfo{person}{Been Kim}.} \bibinfo{year}{2018}\natexlab{}.
\newblock \showarticletitle{Considerations for evaluation and generalization in
  interpretable machine learning}.
\newblock In \bibinfo{booktitle}{\emph{Explainable and interpretable models in
  computer vision and machine learning}}. \bibinfo{publisher}{Springer},
  \bibinfo{pages}{3--17}.
\newblock


\bibitem[\protect\citeauthoryear{Hart}{Hart}{2006}]%
        {hart2006nasa}
\bibfield{author}{\bibinfo{person}{Sandra~G Hart}.}
  \bibinfo{year}{2006}\natexlab{}.
\newblock \showarticletitle{NASA-task load index (NASA-TLX); 20 years later}.
  In \bibinfo{booktitle}{\emph{Proceedings of the human factors and ergonomics
  society annual meeting}}, Vol.~\bibinfo{volume}{50}. Sage publications Sage
  CA: Los Angeles, CA, \bibinfo{pages}{904--908}.
\newblock


\bibitem[\protect\citeauthoryear{Lipton}{Lipton}{2018}]%
        {lipton2018mythos}
\bibfield{author}{\bibinfo{person}{Zachary~C Lipton}.}
  \bibinfo{year}{2018}\natexlab{}.
\newblock \showarticletitle{The Mythos of Model Interpretability: In machine
  learning, the concept of interpretability is both important and slippery.}
\newblock \bibinfo{journal}{\emph{Queue}} \bibinfo{volume}{16},
  \bibinfo{number}{3} (\bibinfo{year}{2018}), \bibinfo{pages}{31--57}.
\newblock


\bibitem[\protect\citeauthoryear{Lundberg and Lee}{Lundberg and Lee}{2017}]%
        {lundberg2017unified}
\bibfield{author}{\bibinfo{person}{Scott~M Lundberg} {and}
  \bibinfo{person}{Su-In Lee}.} \bibinfo{year}{2017}\natexlab{}.
\newblock \showarticletitle{A unified approach to interpreting model
  predictions}.
\newblock \bibinfo{journal}{\emph{Advances in neural information processing
  systems}}  \bibinfo{volume}{30} (\bibinfo{year}{2017}).
\newblock


\bibitem[\protect\citeauthoryear{Macdonald and Tonellotto}{Macdonald and
  Tonellotto}{2020}]%
        {pyterrier2020ictir}
\bibfield{author}{\bibinfo{person}{Craig Macdonald} {and}
  \bibinfo{person}{Nicola Tonellotto}.} \bibinfo{year}{2020}\natexlab{}.
\newblock \showarticletitle{Declarative Experimentation inInformation Retrieval
  using PyTerrier}. In \bibinfo{booktitle}{\emph{Proceedings of ICTIR 2020}}.
\newblock


\bibitem[\protect\citeauthoryear{Nauta, Trienes, Pathak, Nguyen, Peters,
  Schmitt, Schl{\"o}tterer, van Keulen, and Seifert}{Nauta
  et~al\mbox{.}}{2022}]%
        {nauta2022anecdotal}
\bibfield{author}{\bibinfo{person}{Meike Nauta}, \bibinfo{person}{Jan Trienes},
  \bibinfo{person}{Shreyasi Pathak}, \bibinfo{person}{Elisa Nguyen},
  \bibinfo{person}{Michelle Peters}, \bibinfo{person}{Yasmin Schmitt},
  \bibinfo{person}{J{\"o}rg Schl{\"o}tterer}, \bibinfo{person}{Maurice van
  Keulen}, {and} \bibinfo{person}{Christin Seifert}.}
  \bibinfo{year}{2022}\natexlab{}.
\newblock \showarticletitle{From anecdotal evidence to quantitative evaluation
  methods: A systematic review on evaluating explainable ai}.
\newblock \bibinfo{journal}{\emph{arXiv preprint arXiv:2201.08164}}
  (\bibinfo{year}{2022}).
\newblock


\bibitem[\protect\citeauthoryear{Nguyen and Mart{\'\i}nez}{Nguyen and
  Mart{\'\i}nez}{2020}]%
        {nguyen2020quantitative}
\bibfield{author}{\bibinfo{person}{An-phi Nguyen} {and}
  \bibinfo{person}{Mar{\'\i}a~Rodr{\'\i}guez Mart{\'\i}nez}.}
  \bibinfo{year}{2020}\natexlab{}.
\newblock \showarticletitle{On quantitative aspects of model interpretability}.
\newblock \bibinfo{journal}{\emph{arXiv preprint arXiv:2007.07584}}
  (\bibinfo{year}{2020}).
\newblock


\bibitem[\protect\citeauthoryear{Olmsted-Hawala, Murphy, Hawala, and
  Ashenfelter}{Olmsted-Hawala et~al\mbox{.}}{2010}]%
        {olmsted2010think}
\bibfield{author}{\bibinfo{person}{Erica~L Olmsted-Hawala},
  \bibinfo{person}{Elizabeth~D Murphy}, \bibinfo{person}{Sam Hawala}, {and}
  \bibinfo{person}{Kathleen~T Ashenfelter}.} \bibinfo{year}{2010}\natexlab{}.
\newblock \showarticletitle{Think-aloud protocols: a comparison of three
  think-aloud protocols for use in testing data-dissemination web sites for
  usability}. In \bibinfo{booktitle}{\emph{Proceedings of the SIGCHI conference
  on human factors in computing systems}}. \bibinfo{pages}{2381--2390}.
\newblock


\bibitem[\protect\citeauthoryear{Polley, Koparde, Gowri, Perera, and
  Nuernberger}{Polley et~al\mbox{.}}{2021}]%
        {polley2021towards}
\bibfield{author}{\bibinfo{person}{Sayantan Polley},
  \bibinfo{person}{Rashmi~Raju Koparde}, \bibinfo{person}{Akshaya~Bindu Gowri},
  \bibinfo{person}{Maneendra Perera}, {and} \bibinfo{person}{Andreas
  Nuernberger}.} \bibinfo{year}{2021}\natexlab{}.
\newblock \showarticletitle{Towards trustworthiness in the context of
  explainable search}. In \bibinfo{booktitle}{\emph{Proceedings of the 44th
  International ACM SIGIR Conference on Research and Development in Information
  Retrieval}}. \bibinfo{pages}{2580--2584}.
\newblock


\bibitem[\protect\citeauthoryear{Qu, Arguello, and Wang}{Qu
  et~al\mbox{.}}{2020}]%
        {qu2020towards}
\bibfield{author}{\bibinfo{person}{Jiaming Qu}, \bibinfo{person}{Jaime
  Arguello}, {and} \bibinfo{person}{Yue Wang}.}
  \bibinfo{year}{2020}\natexlab{}.
\newblock \showarticletitle{Towards explainable retrieval models for precision
  medicine literature search}. In \bibinfo{booktitle}{\emph{Proceedings of the
  43rd International ACM SIGIR Conference on Research and Development in
  Information Retrieval}}. \bibinfo{pages}{1593--1596}.
\newblock


\bibitem[\protect\citeauthoryear{Ramos and Eickhoff}{Ramos and
  Eickhoff}{2020}]%
        {ramos2020search}
\bibfield{author}{\bibinfo{person}{Jerome Ramos} {and} \bibinfo{person}{Carsten
  Eickhoff}.} \bibinfo{year}{2020}\natexlab{}.
\newblock \showarticletitle{Search result explanations improve efficiency and
  trust}. In \bibinfo{booktitle}{\emph{Proceedings of the 43rd International
  ACM SIGIR Conference on Research and Development in Information Retrieval}}.
  \bibinfo{pages}{1597--1600}.
\newblock


\bibitem[\protect\citeauthoryear{Ribeiro, Singh, and Guestrin}{Ribeiro
  et~al\mbox{.}}{2016}]%
        {ribeiro2016should}
\bibfield{author}{\bibinfo{person}{Marco~Tulio Ribeiro},
  \bibinfo{person}{Sameer Singh}, {and} \bibinfo{person}{Carlos Guestrin}.}
  \bibinfo{year}{2016}\natexlab{}.
\newblock \showarticletitle{" Why should i trust you?" Explaining the
  predictions of any classifier}. In \bibinfo{booktitle}{\emph{Proceedings of
  the 22nd ACM SIGKDD international conference on knowledge discovery and data
  mining}}. \bibinfo{pages}{1135--1144}.
\newblock


\bibitem[\protect\citeauthoryear{Singh and Anand}{Singh and Anand}{2019}]%
        {singh2019exs}
\bibfield{author}{\bibinfo{person}{Jaspreet Singh} {and}
  \bibinfo{person}{Avishek Anand}.} \bibinfo{year}{2019}\natexlab{}.
\newblock \showarticletitle{Exs: Explainable search using local model agnostic
  interpretability}. In \bibinfo{booktitle}{\emph{Proceedings of the Twelfth
  ACM International Conference on Web Search and Data Mining}}.
  \bibinfo{pages}{770--773}.
\newblock


\bibitem[\protect\citeauthoryear{Voorhees}{Voorhees}{2005}]%
        {voorhees_overview_2005}
\bibfield{author}{\bibinfo{person}{Ellen Voorhees}.}
  \bibinfo{year}{2005}\natexlab{}.
\newblock \bibinfo{title}{Overview of the {TREC} 2004 {Robust} {Retrieval}
  {Track}}.
\newblock
\newblock
\urldef\tempurl%
\url{https://doi.org/10.6028/NIST.SP.500-261}
\showDOI{\tempurl}


\bibitem[\protect\citeauthoryear{Yu, Rahimi, and Allan}{Yu
  et~al\mbox{.}}{2022}]%
        {yu2022towards}
\bibfield{author}{\bibinfo{person}{Puxuan Yu}, \bibinfo{person}{Razieh Rahimi},
  {and} \bibinfo{person}{James Allan}.} \bibinfo{year}{2022}\natexlab{}.
\newblock \showarticletitle{Towards Explainable Search Results: A Listwise
  Explanation Generator}. In \bibinfo{booktitle}{\emph{Proceedings of the 45th
  International ACM SIGIR Conference on Research and Development in Information
  Retrieval}}. \bibinfo{pages}{669--680}.
\newblock


\end{thebibliography}

\end{document}